\documentclass[journal, comsoc]{IEEEtran}
\usepackage{hyperref}
\usepackage{booktabs}
\usepackage{makecell}
\IEEEoverridecommandlockouts
\usepackage{cite}
\usepackage{amsmath,amssymb,amsfonts}
\usepackage{algorithmic}
\usepackage{algorithm}
\usepackage{graphicx}
\usepackage{textcomp}
\usepackage{xcolor}
\usepackage[caption=false,font=footnotesize,
labelfont=rm,textfont=rm]{subfig}
\renewcommand{\baselinestretch}{0.962}

\begin{document}
\title{A General Optimization Framework for Movable Antenna Systems via Discrete Sampling}

\author{Changhao Liu, Weidong Mei, \IEEEmembership{Member, IEEE}, Zhi Chen, \IEEEmembership{Senior Member, IEEE}, Jun Fang, \IEEEmembership{Senior Member, IEEE}, Boyu Ning, \IEEEmembership{Member, IEEE}\vspace{-6pt}
\thanks{
The authors are with the National Key Laboratory of Wireless Communications, University of Electronic Science and Technology of China, Chengdu 611731, China (e-mail: lch3001@163.com, wmei@uestc.edu.cn, chenzhi@uestc.edu.cn, junfang@uestc.edu.cn, boydning@outlook.com).
}
}
\maketitle

\begin{abstract}
Movable antenna (MA) systems have attracted growing interest in wireless communications due to their ability to reshape wireless channels via local antenna movement within a confined region. However, optimizing antenna positions to enhance communication performance turns out to be challenging due to the highly nonlinear relationship between wireless channels and antenna positions. Existing approaches, such as gradient-based and heuristic algorithms, often suffer from high computational complexity or undesired local optima. To address the above challenge, this letter proposes a general and low-complexity optimization framework for MA position optimization. Specifically, we discretize the antenna movement region into a set of sampling points, thereby transforming the continuous optimization problem into a discrete point selection problem. Next, we sequentially update the optimal sampling point for each MA over multiple rounds. To avoid convergence to poor local optima, a Gibbs sampling (GS) phase is introduced between rounds to explore adjacent and randomly generated candidate solutions. As a case study, we investigate joint precoding and antenna position optimization for an MA-enhanced broadcast system by applying the proposed framework. Numerical results demonstrate that the proposed algorithm achieves near-optimal performance and significantly outperforms existing benchmarks.\vspace{-9pt}
\end{abstract}
\begin{IEEEkeywords}
Movable antennas (MAs), antenna position optimization, discrete sampling, sequential update, Gibbs sampling, broadcast system.
\end{IEEEkeywords}\vspace{-6pt}

\section{Introduction}
Movable antenna (MA) technology has recently attracted significant attention from both academia and industry, as it enables flexible repositioning of multiple antennas within a confined region. Thanks to this new degree of freedom (DoF), MAs are expected to achieve comparable or even superior performance to conventional fixed-position antennas (FPAs) with significantly fewer radio-frequency (RF) chains \cite{zhu2025tutorial}. Prior works have demonstrated that MAs can offer a broad range of advantages over FPAs, such as desired signal enhancement\cite{zhu2024modeling, mei2024movable,zeng2025csi}, interference suppression\cite{xiao2024multiuser, zhu2024multiuser, wei2024spectrum, wei2025movable}, flexible array signal processing\cite{zhu2023movable, wang2025movable}, improved sensing accuracy\cite{lyu2025movable,shao2025exploit}, among others.

However, realizing the above benefits of MAs hinges on effective antenna position optimization, which is inherently challenging due to the highly nonlinear relationship between antenna positions and wireless channels. While various algorithms have been proposed to solve this problem, each of them has certain limitations. A common approach is to use gradient-based methods such as successive convex approximation (SCA) and gradient descent\cite{zhu2024multiuser, wang2025movable, hu2024secure}. However, these methods are often prone to getting trapped in poor local optima. Moreover, gradient computation becomes increasingly challenging in the presence of complex objective functions or constraints, as commonly encountered in MA systems. To address these difficulties, heuristic algorithms like particle swarm optimization (PSO) and evolutionary algorithms have also been explored \cite{ding2025movable, xiao2024multiuser}. Yet, these methods lack even local optimality guarantees and tend to incur high computational complexity.
More recently, some works have proposed discretizing the antenna movement region into a set of sampling points, thereby converting the continuous optimization problem into a discrete point selection problem with inter-antenna spacing constraints. For example, the authors in \cite{mei2024movable,ma2025robust} showed that for a single-user multiple-input single-output (MISO) system, the optimal MA positions can be efficiently obtained using a graph-based algorithm in polynomial time. While for other more general scenarios, some studies have proposed combining discrete point selection with solution pruning to obtain the optimal MA positions\cite{mei2024secure,wu2025globally}; however, its worst-case computational complexity is exponential. To the best of our knowledge, a general MA position optimization framework that strikes a practical balance between performance and complexity is still lacking.

\begin{figure}[tb]
\centerline{\includegraphics[width=0.33\textwidth]{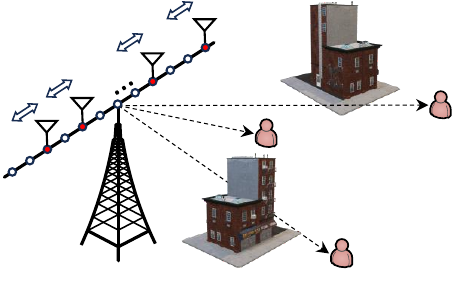}}
\vspace{-15pt}
\caption{MA-enhanced broadcast system.}
\label{sysmod}
\vspace{-15pt}
\end{figure}
To tackle this challenge, we propose a general and low-complexity optimization framework for MA position optimization in this letter, inspired by the discretization approach adopted in \cite{mei2024movable,ma2025robust,mei2024secure,wu2025globally}. Unlike prior works that rely on solution pruning to select optimal sampling points, we iteratively update the sampling point for each MA over multiple rounds. To escape low-quality local optima, a Gibbs sampling (GS) phase is introduced between two consecutive rounds to explore adjacent and randomly generated candidate solutions. The proposed algorithm is shown to only incur a linear complexity, which is dramatically lower than existing algorithms with polynomial or exponential complexity. As a case study, we apply this framework to jointly optimize precoding and MA positions in an MA-enhanced broadcast system (see Fig.~\ref{sysmod}). Numerical results demonstrate that the proposed algorithm achieves near-optimal performance and significantly outperforms existing benchmarks.

\begingroup
\allowdisplaybreaks
\section{System Model and Problem Formulation}
We consider the downlink transmission from a multi-antenna BS to $K$ users. Note that the proposed algorithm is also applicable to the uplink scenario. The BS is assumed to be equipped with $N$ MAs that can be flexibly moved within a transmit region, denoted as $\mathcal{C}_t$. 
Denote by $\mathbf{w}_k \in \mathbb{C}^{N \times 1}, \;k \in \mathcal{K} \triangleq \{1,2,\cdots,K\}$ and $x_n \in \mathcal{C}_t, \;n \in \mathcal{N} \triangleq \{1,2,\cdots,N\}$ the transmit beamforming vector for user $k$ and the position of the $n$-th MA at the BS, respectively. 
In this letter, we aim to jointly optimize the BS's transmit precoding matrix $\mathbf{W}=[\mathbf{w}_1, \mathbf{w}_2,\cdots,\mathbf{w}_K] \in \mathbb{C}^{N \times K}$ and antenna position vector $\mathbf{x}=[x_1,x_2,\cdots,x_N]^T$. A general optimization problem is formulated as
\begin{align} 
(\mathrm{P}1) \max_{{\bf x},\mathbf{W} } \;\;\;&U({\bf x},\mathbf{W}) \notag\\ 
\mathrm{s.t.}\;\;\; &f_i({\bf x}) \ge 0, \;\;\;\;\;\;1 \le i \le I_1, \label{st1}\\
&g_i(\mathbf{W}) \ge 0, \;\;\;\;1 \le i \le I_2, \label{st2}\\
&q_i({\bf x,W}) \ge 0, \,1 \le i \le I_3,\label{st3}
\end{align}
where $U(\cdot)$ denotes a prescribed utility function, e.g., multi-user sum rate, multicast rate, or minimum signal-to-interference-plus-noise ratio (SINR) among all users; $f_i(\cdot)$ denotes the constraints associated with the MAs, e.g., the minimum inter-MA spacing and finite movable region; $g_i(\cdot)$ represents the constraints associated with the BS's transmit precoding, e.g., maximum transmit power; and $q_i(\cdot)$ represents the coupled constraints involving both antenna positions and transmit precoding. Moreover, $I_1$, $I_2$, and $I_3$ denote the number of constraints for each of the above types.

For ease of exposition, we recast (P1) as a simplified problem (P2) with respect to (w.r.t.) the MA position vector $\mathbf{x}$ only. This corresponds to two scenarios for optimizing (P1). In the first scenario, alternating optimization (AO) is applied to solve (P1), where $\mathbf{x}$ and $\mathbf{W}$ are alternately optimized with the other being fixed until convergence, and we focus on the subproblem of optimizing $\mathbf{x}$ for any given $\mathbf{W}$. In the second scenario, we have derived an optimal $\mathbf{W}$ for (P1) in terms of $\mathbf{x}$ (e.g., maximum ratio transmission (MRT) in the single-user scenario), denoted as $\mathbf{W}(\mathbf{x})$, and it suffices to optimize $\mathbf{x}$ only. We take the second scenario as an example to present (P2), i.e.,
\begin{equation}
(\mathrm{P}2)\; \max_{{\bf x}} \;U({\bf x},\mathbf{W}(\mathbf{x})), \;\;
\mathrm{s.t.}\;\;{\text{\eqref{st1}}},
\end{equation}
where the constraints in \eqref{st2} and \eqref{st3} are removed, as they have been accounted for in deriving $\mathbf{W}(\mathbf{x})$ for any given $\mathbf{x}$.\footnote{In the first scenario with AO, only the constraints in \eqref{st2} should be removed by treating $\mathbf{W}$ as a constant.}

Although (P2) is a simplified version of (P1), it is still difficult to be optimally solved. The reason is that the wireless channel is generally a highly non-convex function in terms of the MA position vector $\mathbf{x}$. Moreover, the complex expression of $\mathbf{W}(\mathbf{x})$ and the continuous nature of $\mathbf{x}$ render the objective functions and constraints in (P2) even more challenging to tackle. To solve (P2), we propose a general optimization framework for MAs based on discrete sampling, as presented next.
\vspace{-4pt}

\section{Proposed General Optimization Framework}\label{PGOF}
In our proposed framework, we first discretize the movement region into a multitude of discrete points and then employ the sequential update algorithm and GS method to obtain high-quality local optimal MA positions.\vspace{-6pt}
\vspace{-3pt}

\subsection{Discrete Sampling and Sequential Update}\label{PGOF-a}
Specifically, we sample the transmit region ${\cal C}_t$ into $M$ discrete points uniformly ($M \gg N$). Notably, as $M$ is sufficiently large, each MA can be approximately located at one of the $M$ sampling points. Assume that the $n$-th MA is located at the $a_n$-th sampling point, with $a_n \in \mathcal{M} \triangleq \{1, 2, \cdots, M\}$. Hence, in the case of a linear antenna array, the position of the $n$-th MA is given by $x_n = \frac{a_n A}{M}$, where $A$ denotes the length of the antenna array. Moreover, define $\mathbf{a} = [a_1, a_2, \cdots, a_N]^T$ as the index vector of all $N$ MAs. As such, (P2) can be converted into a discrete optimization problem by replacing $x_n$ therein with $\frac{a_nA}{M}$, denoted as (P3).
\begin{equation}
(\mathrm{P}3) \max_{{\bf a}} \;U({\bf a},\mathbf{W}(\mathbf{a})), \;\; 
\mathrm{s.t.}\;\;\; f_i({\bf a}) \ge 0, \;\;1 \le i \le I_1. \label{st33}
\end{equation}

\textcolor{blue}{Notably, the formulation of (P3) requires only point-wise channel state information (CSI). In the presence of imperfect CSI, (P3) can be extended into a robust position optimization problem.}
To solve (P3), we first employ the sequential update algorithm, which sequentially searches the best position of each MA within $\mathcal{M}$ in multiple rounds. Denote by $\mathbf{a}^{(l-1)} = [a_1^{(l-1)}, a_2^{(l-1)}, \cdots, a_N^{(l-1)}]^T$ the index set of the $N$ MAs after the $(l-1)$-th round of the sequential search. In the $n$-th iteration of the $l$-th round of the sequential update, we update the index of the $n$-th MA and denote by $a^{\star}_{n,l}$ its updated index. It follows that $a^{\star}_{n,l}$ can be obtained as
\begin{equation}
a_{n,l}^\star = \arg \max_{a_n \in \Psi_{n,l}} U\Big(\tilde{\mathbf{a}}_l(a_n), \mathbf{W}\big(\tilde{\mathbf{a}}_l(a_n)\big)\Big), n \in \mathcal{N},
\end{equation}
where
\begin{align}
\tilde{\mathbf{a}}_l&(a_n) = \notag\\
&\begin{cases}
[a_{1}, a_2^{(l-1)}, \cdots, a_N^{(l-1)}]^T,  &n=1\\
[a_{1,l}^{\star},\cdots, a_{n-1,l}^{\star}, a_n, a_{n+1}^{(l-1)}, \cdots, a_N^{(l-1)}]^T,  &2 \le n \le N\!-\!1\\
[a_{1,l}^{\star}, \cdots, a_{N-1,l}^{\star}, a_N]^T,&n=N,
\end{cases}\notag
\end{align}
and $\Psi_{n,l}$ denotes the feasibility set of $a_n$ given $\mathbf{a}=\tilde{\mathbf{a}}_l(a_n)$ in (P3).

After $a_{N,l}^\star$ is obtained, we update $a_n^{(l)} = a^{\star}_{n,l}, n \in \mathcal{N}$, and the optimized MA index vector after the $l$-th round of sequential update is given by 
\begin{equation}
\mathbf{a}^{(l)} = [a_1^{(l)}, a_2^{(l)}, \cdots, a_N^{(l)}]^T=[a_{1,l}^{\star}, a_{2,l}^{\star}, \cdots, a_{N,l}^{\star}]^T. \label{asu}
\end{equation}
However, the optimized MA positions in \eqref{asu} may not be optimal, as the sequential update may be trapped by low-quality local optimal solutions. To further improve its performance, we propose an additional GS phase to explore more feasible solutions to (P3), thereby circumventing local optimality.\vspace{-6pt}

\subsection{Gibbs Sampling Phase}
GS takes place between two consecutive rounds of the sequential update. After the $l$-th round of the sequential update, the GS takes $\mathbf{a}^{(l)}$ in \eqref{asu} as its input. 
Consider the $t$-th GS iteration and let $\mathcal{E}(t-1) = \{\mathbf{a}_{\text{GS}}^{(0)}, \mathbf{a}^{(1)}_{\text{GS}}, \cdots, \mathbf{a}^{(t-1)}_{\text{GS}}\}$ denote the optimized index vectors by the previous GS iterations, where $\mathbf{a}^{(i)}_{\text{GS}} = [a^{(i)}_{\text{GS},1}, a^{(i)}_{\text{GS},2}, \cdots, a^{{(i)}}_{\text{GS},N}]^T$ denotes the output of the $i$-th GS iteration, with $a^{{(i)}}_{\text{GS},n}$ denoting the index of the $n$-th MA in this iteration. In particular, we set $\mathbf{a}_{\text{GS}}^{(0)} = \mathbf{a}^{(l)}$.

At the beginning of the $t$-th GS iteration, we generate $S$ candidate solutions for exploration, including $S^{\text{adj}}$ adjacent solutions to $\mathbf{a}^{(t-1)}_{\text{GS}}$ and $(S - S^{\text{adj}})$ random solutions. Each adjacent solution is obtained by shifting the index of a specific MA by a certain amount (with those of other MAs fixed). Let $J$ denote the maximum index shift. As a result, we can obtain the following $2NJ$ adjacent solutions to $\mathbf{a}^{(t-1)}_{\text{GS}}$, i.e., 
\begin{align} 
\mathbf{a}^{\!(t,-j)}_{n,{\text{adj}}} &= [a^{\!(t\!-\!1)}_{\text{GS},1}, \cdots,  {a^{\!(t\!-\!1)}_{\text{GS},n}\!-\!j}, \cdots, a^{\!(t\!-\!1)}_{\text{GS},N}]^T, n \in \mathcal{N}, j \in \mathcal{J}, \label{aadj1}\\[-3pt]
\mathbf{a}^{\!(t,j)}_{n,{\text{adj}}} &= [a^{\!(t\!-\!1)}_{\text{GS},1}, \cdots,  a^{\!(t\!-\!1)}_{\text{GS},n}\!+\!j, \cdots, a^{\!(t\!-\!1)}_{\text{GS},N}]^T, n \in \mathcal{N}, j \in \mathcal{J}, \notag%\label{aadj2}
\end{align}
where $\mathcal{J}\triangleq \{1, 2, \cdots, J\}$. Note that some of these $2NJ$ solutions may not satisfy the constraints in (P3). Hence, we only choose $S^{\text{adj}}$ feasible adjacent solutions from them, with $S^{\text{adj}} \le 2NJ$. Let $\mathcal{B}(t)$ denote the set of adjacent solutions in the $t$-th GS iteration. Next, we randomly generate $(S - S^{\text{adj}})$ feasible solutions to (P3) and denote their set as $\mathcal{D}(t)$. 
The details on how to generate these random solutions will be provided in the next section based on a concrete example. Let $\bar{\mathbf{a}}^{(t)}_{\text{GS},s}$ be the $s$-th candidate solution in the set $\mathcal{B}(t) \cup \mathcal{D}(t)$.

Next, we select one solution as the output of the $t$-th GS iteration, i.e.,  $\mathbf{a}^{(t)}_{\text{GS}}$, from the $S$ candidate solutions in $\mathcal{B}(t) \cup \mathcal{D}(t)$ given $\mathbf{a}_{\text{GS}}^{(t-1)}$. The probability that the $s$-th candidate solution in $\mathcal{B}(t) \cup \mathcal{D}(t)$, $1 \le s \le S$, is selected is given by\cite{bremaud2013markov}
\begin{equation}
P_s^{(t)}={\text{Pr}}\big\{\mathbf{a}_{\text{GS}}^{(t)}=\bar{\mathbf{a}}^{(t)}_{\text{GS},s}|\mathbf{a}_{\text{GS}}^{(t-1)}\big\} =\frac{e^{\mu U\big(\bar{\mathbf{a}}^{(t)}_{\text{GS},s},\mathbf{W}(\bar{\mathbf{a}}^{(t)}_{\text{GS},s})\big)}}{\sum\nolimits_{\mathbf{a}\in \mathcal{B}(t) \cup \mathcal{D}(t)} e^{\mu U\big(\mathbf{a},\mathbf{W}(\mathbf{a})\big) }}, \label{pr}
\end{equation} 
where $\mu > 0$ is a pre-defined scaling parameter. 
To determine $\mathbf{a}_{\text{GS}}^{(t)}$ based on \eqref{pr}, we randomly generate a value between 0 and 1, denoted as $p_t$, and determine $\mathbf{a}_{\text{GS}}^{(t)}$ as
\begin{equation}
\mathbf{a}_{\text{GS}}^{(t)} = \bar{\mathbf{a}}^{(t)}_{\text{GS},s^{\star}} \label{at}
\end{equation}
where $s^\star$ is the index satisfying $\sum_{s=1}^{s^\star-1}P_s^{(t)} < p_t \le \sum_{s=1}^{s^\star}P_s^{(t)}$. Then, we update $\mathcal{E}(t) = \mathbf{a}_{\text{GS}}^{(t)} \cup \mathcal{E}(t-1)$ to finalize the $t$-th GS iteration.

The GS proceeds until the iteration number $t$ reaches a pre-defined maximum number, denoted by $T$. Finally, among all solutions in $\mathcal{E}(T)$, we modify \eqref{asu} as the solution that yields the maximum value of $U(\cdot)$, i.e., 
\begin{equation}
\mathbf{a}^{(l)} = \arg \max_{\mathbf{a} \in \mathcal{E}(T)} U\big(\mathbf{a},\mathbf{W}(\mathbf{a})\big), \label{ags} 
\end{equation}
and the $(l+1)$-th round of the sequential update follows. We summarize the proposed general framework in Algorithm \ref{alg1}. 

Note that as $\mathcal{E}(t-1)$ includes the output of the sequential update, i.e., $\mathbf{a}^{(l)}$, it always yields a performance no worse than the $l$-th round of the sequential update. Since the sequential update algorithm outputs a non-decreasing objective value of (P3), Algorithm \ref{alg1} is ensured to converge. \textcolor{blue}{Moreover, the complexity of the sequential update is no more than $\mathcal{O}(KLMN)$, while that of the GS phase is given by $\mathcal{O}(KST)$}. As such, Algorithm \ref{alg1} yields a linear complexity order in terms of $M$ and $N$, which is much lower than those of existing gradient-based and heuristic algorithms with polynomial complexity orders. Moreover, compared to the gradient-based algorithm, the proposed algorithm dispenses with complex gradient calculation and can be applied to any problem structure. Whereas compared to the heuristic algorithms, the proposed algorithm ensures high-quality local optimality thanks to the GS.
\vspace{-6pt}

\begin{algorithm}[t] 
\caption{Proposed Optimization Framework} \label{alg1}
\begin{algorithmic}
\STATE Input: $\mathbf{a}^{(0)}$, $L$, $T$
\STATE $l \leftarrow 1$, $t \leftarrow 1$
\STATE \textbf{while} $l<L$:
\STATE \hspace{0.5cm} Calculate $\mathbf{a}^{(l)}$ based on \eqref{asu} via sequential update.
\STATE \hspace{0.5cm} \textbf{while} $t<T$:
\STATE \hspace{0.5cm} \hspace{0.5cm} Generate $\mathcal{B}(t)$ via \eqref{aadj1}.
\STATE \hspace{0.5cm} \hspace{0.5cm} Generate $\mathcal{D}(t)$.
\STATE \hspace{0.5cm} \hspace{0.5cm} Determine $\mathbf{a}_{\text{GS}}^{(t)}$ based on \eqref{pr} and \eqref{at}.
\STATE \hspace{0.5cm} \hspace{0.5cm} Update $\mathcal{E}(t) = \mathcal{E}(t-1) \cup \mathbf{a}_{\text{GS}}^{(t)}$.
\STATE \hspace{0.5cm} \hspace{0.5cm} $t \leftarrow t+1$
\STATE \hspace{0.5cm} \textbf{end while}
\STATE \hspace{0.5cm} Update $\mathbf{a}^{(l)}$ based on \eqref{ags}.
\STATE \hspace{0.5cm} $l \leftarrow l+1$
\STATE \textbf{end while}
\STATE Output: $\mathbf{a}^{(L)}$
\end{algorithmic}
\end{algorithm}

\section{Case Study}
In this section, we consider an MA-enhanced broadcast system as a case study to implement the proposed algorithm, as shown in Fig.~\ref{sysmod}, where a multi-MA BS transmits to multiple users each equipped with a single FPA. The MAs can be flexibly moved within a linear array of length $A$.
To avoid mutual coupling, we set a minimum distance $d_{\min}$ between any two adjacent MAs. As such, given the discrete sampling in Section \ref{PGOF-a}, it must hold that 
\begin{equation} 
A{|a_i - a_j|}/{M} \ge d_{\min}, \;\; i \neq j,\;\; i,j \in \mathcal{N}, \label{dmin}
\end{equation}
which leads to $|a_i - a_j|\ge a_{\min}$, with $a_{\min} = {d_{\min}M}/{A}$ (assumed to be an integer).
Denote by $h_{m}^{(k)} \in \mathbb{C}, m \in \mathcal{M}, k \in \mathcal{K}$ the baseband equivalent channel from the $m$-th discrete sampling point to user $k$, which is assumed to be already known by the BS via various channel estimation techniques for MAs \cite{zhu2025tutorial}. Hence, the channel from the BS to user $k$ is expressed as
\begin{equation} \label{hk}
\mathbf{h}_{k}(\mathbf{a}) = [h_{a_1}^{(k)}, h_{a_2}^{(k)}, \cdots, h_{a_N}^{(k)}]^H,\; k \in \mathcal{K}.
\end{equation}
The received signal at user $k$ can be expressed as
\begin{align}	
y_k(\mathbf{a},\mathbf{W}) &= \mathbf{h}_{k}^H(\mathbf{a})\mathbf{W}\mathbf{s} + n_k, \notag \\
&= \mathbf{h}_{k}^H(\mathbf{a}) \Big( \mathbf{w}_k s_k + \sum\nolimits_{i \neq k}\mathbf{w}_i s_i \Big) + n_k, \;k \in \mathcal{K}, \label{y_k}
\end{align}
where $\mathbf{s} = [s_1, s_2, \cdots, s_K]^H $ denotes the transmit symbols for the $K$ users with $\mathbb{E}(\mathbf{s}\mathbf{s}^H) = \mathbf{I}_K$, and $n_k \sim {\cal{CN}}(0, \sigma^2)$ is the additive white Gaussian noise (AWGN) at user $k$ with $\sigma^2$ denoting the noise power. Hence, the achievable rate at user $k$ is given by
%\vspace{-5pt}
\begin{equation}	\label{rk} 
R_k\big(\mathbf{a},\mathbf{W}\big) = \log_2 \Big( 1 + \frac{|\mathbf{h}^H_k(\mathbf{a}) \mathbf{w}_k|^2}
{\sum_{i \neq k}|\mathbf{h}^H_k(\mathbf{a}) \mathbf{w}_i|^2 + \sigma^2} \Big), k \in \mathcal{K}, 
\end{equation}

For any given $\mathbf{a}$, we apply a regularized zero forcing (RZF) strategy to design the precoding matrix $\mathbf{W}$ at the BS for simplicity, which is given by
\begin{equation}
\mathbf{W}_{\text{RZF}}(\mathbf{a},\rho) = (\mathbf{H}(\mathbf{a})\mathbf{H}^H(\mathbf{a}) + \rho \mathbf{I}_N)^{-1}\mathbf{H}(\mathbf{a}), \label{wstar}
\end{equation}
where $\mathbf{H}(\mathbf{a}) = [\mathbf{h}_1(\mathbf{a}), \mathbf{h}_2(\mathbf{a}), \cdots, \mathbf{h}_K(\mathbf{a})]^H$ denotes the channel matrix between the BS and all users, and  $\rho > 0$ denotes a tunable parameter. Notably, by properly setting the parameter $\rho$, the RZF precoding can achieve a close performance to the optimal precoding design.

The goal of this letter is to maximize the sum rate of all users by optimizing the BS's MA index vector $\mathbf{a}$ and the RZF parameter $\rho$. Hence, the associated optimization problem is formulated as
\vspace{-5pt}
\begin{align}	
(\mathrm{P}4)\; \max_{\mathbf{a}, \rho}\;\;&\sum_{k=1}^{K}R_k\big(\mathbf{a},\mathbf{W}_{\text{RZF}}(\mathbf{a},\rho)\big) \notag \\  
\mathrm{s.t.} \;
\;&|a_i - a_j| \ge a_{\min}, \; i \neq j,\; i,j \in \mathcal{N}, \label{p3c1} \\ 
\;&\|\mathbf{W}_{\text{RZF}}(\mathbf{a},\rho)\|_2^2 \le P, \label{p3c2} 
\end{align} 
where $P$ denotes the maximum transmit power of the BS. Note that for any given $\mathbf{a}$, (P4) is a single-variable optimization problem, for which the optimal $\rho$, denoted as $\rho^\star(\mathbf{a})$, can be numerically obtained via a bisection search, since $\|\mathbf{W}_{\text{RZF}}(\mathbf{a},\rho)\|_2^2$ is a non-decreasing function of $\rho$. As such, we can simplify (P4) as an optimization problem with respect to (w.r.t.) $\bf a$ only, labeled as (P5), by replacing $\rho$ therein with $\rho^\star(\mathbf{a})$ and removing \eqref{p3c2}. For (P5), we can employ the proposed general framework in Section \ref{PGOF}.

Specifically, in the sequential update process, the feasibility set of $a_n$ can be explicitly expressed as
\begin{align}
\Psi_{n,l} = \big\{m\in\mathcal{M}&\big|\lvert m-a_{i,l}^\star \rvert \ge a_{\min}, 1 \le i \le n-1, 
\big \lvert{m-a_j^{(l-1)}}\big \rvert  \notag \\
&\ge a_{\min}, n+1\le j\le N\big\},\; 2\!\le \!n\! \le\! N-1,
\end{align}
with \textcolor{blue}{$\Psi_{1,l} = \{m|m\in\mathcal{M}, \lvert m-a_j^{(l-1)}\rvert \ge a_{\min}, 2\le j\le N\}$,} and $\Psi_{N,l} = \{m|m\in\mathcal{M}, \lvert m-a_{i,l}^\star \rvert \ge a_{\min}, 1 \le i \le N-1\}$.

Whereas in the GS phase, to generate random solutions subject to \eqref{p3c1}, we can first generate $N$ non-duplicate integers between 1 and $M - (N-1)(a_{\min}-1)$. Then, we add the $n$-th smallest integer among them by $(n-1)(a_{\min}-1)$, such that the minimum spacing between any two antenna indices is satisfied. Mathematically, each randomly generated antenna index vector can be expressed as
\begin{align}
\bar{\mathbf{a}}^{(t)}_{\text{GS},s}& = {\text{sample}}\big(M - (N-1)(a_{\min}-1), N\big) \;+ \notag \\
&[0, a_{\min}-1, \cdots, (N-1)(a_{\min}-1)]^T, S^{\text{adj}}+1 \le s \le S, \notag
\end{align}
where ${\text{sample}}(a,b)$ generates a vector containing $b$ non-duplicate integers from 1 to $a$ in ascending order, e.g. ${\text{sample}}(5,3) = [1,3,4]^T$, and its corresponding solution is given by $[1,5,8]^T$ for $a_{\min} = 3$. 
\textcolor{blue}{It is worth noting that under additional constraints, extra steps may be required to check the feasibility of each randomly generated solution. Infeasible solutions should be discarded, with new ones generated in their places. Furthermore, for a two-dimensional (2D) MA array, random antenna indices can be generated separately for each dimension, followed by the same feasibility checks as described above}. 
\vspace{-6pt}

\section{Numerical Results}
In this section, we present numerical results to evaluate the performance of the proposed optimization framework. Unless otherwise stated, the simulation parameters are set as follows. The number of the BS's transmit MAs is $N=8$. The carrier frequency is 5 GHz, with its wavelength $\lambda = 0.06$ meter (m). The minimum spacing between any two adjacent MAs is set as $d_{\min} = \lambda/2$. 
In the proposed optimization framework, the number of sampling points is $M=48$, and the length of the transmit region is $A=6\lambda$. The number of candidate solutions in each GS iteration is set as $S=3N$, and the maximum number of iterations in each GS phase is set as $T=M$. The maximum index shift in generating the adjacent candidate solutions is set to $J=1$. 
%In the proposed optimization framework, we set the parameters as $M=48$, $A=6\lambda$, $S=3N$, $T=M$ , and $J=1$.
To generate the channels for the sampling points, we adopt the field-response channel model proposed in \cite{zhu2024modeling}, with the number of transmit paths set to $L_t = 9$ for each user. \textcolor{blue}{Note that other more general channel models can also be adopted, as long as point-wise CSI is available.} Let $\gamma_{l,k}$ denote the coefficient for the $l$-th BS-user $k$ transmit path, with $\gamma_{l,k} \sim \mathcal{CN}(0, \beta D_k^{-\alpha}/L_t), l\in\{1,2,\cdots,L_t\}$, where $D_k$ denotes the BS-user $k$ distance, and $\alpha$ and $\beta$ denote the path-loss exponent and the reference path gain, respectively. We set $\alpha = 2.8$ and $\beta=-46$ dB. The angle of departure for each path follows a uniform distribution over $[0,\pi]$. We consider the following three benchmarks for performance comparison.
\begin{itemize}
\item \textbf{Sequential update (SU)}: The MA index vector is optimized via the sequential update algorithm without GS.
\item \textbf{Particle swarm optimization (PSO)}: The MA positions are optimized via the PSO algorithm proposed in \cite{xiao2024multiuser}.
\item \textbf{FPA}: The MAs are fixed at the center of the transmit region with half-wavelength spacing.
\end{itemize}
In the following, we consider two cases, i.e., $K=1$ and $K>1$. All results are averaged over 1000 independent channel realizations.\vspace{-6pt}

\subsection{Single-User Case}
In the single-user case, we set the maximum number of rounds for sequential update as $L=2$. The BS-user distance is set as 100 m.
Note that for any given $\mathbf{a}$, the optimal transmit beamforming in the single-user case can be obtained as the MRT, i.e., $\mathbf{w}(\mathbf{a}) = \sqrt{P}\frac{\mathbf{h}_1(\mathbf{a})}{\left\|\mathbf{h}_1(\mathbf{a})\right\|}$, taking user 1 as an example. For the remaining antenna position optimization, we can obtain its optimal solution by invoking the graph-based algorithm proposed in \cite{mei2024movable}, which can be used to evaluate the optimality gap of the proposed optimization framework.

As shown in Fig.~\ref{sin_use}, we plot the received signal-to-noise ratio (SNR) at the single user versus the number of sampling points $M$. It is observed that the proposed algorithm outperforms the PSO and FPA benchmarks, and its performance gap with the optimal graph-based algorithm is negligible over the whole range of $M$ considered. Moreover, the proposed algorithm yields a better performance than the sequential update algorithm without GS. These observations demonstrate the necessity of the GS and the near-optimality of our proposed algorithm.
\vspace{-6pt}

\begin{figure}[t]
\centerline{\includegraphics[width=0.4\textwidth]{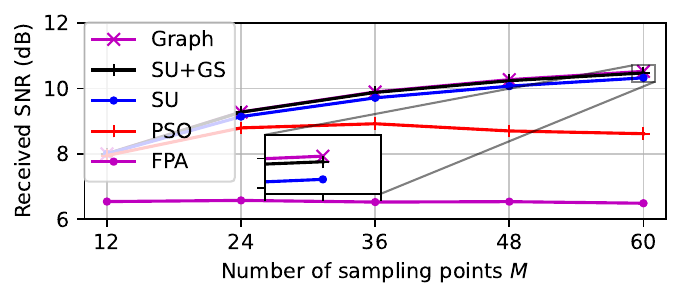}}
\vspace{-12pt}
\caption{Received SNR versus number of sampling points in single-user case.}
\label{sin_use}
\vspace{-12pt}
\end{figure}

\subsection{Multi-User Case} 
In the multi-user case, we set the maximum number of rounds for the sequential update as $L=5$. The number of users is set as $K=3$, with their distances with the BS set as 100, 60, and 40 m, respectively.

In Fig.~\ref{mul_use_m}, we plot the sum rate of all users versus the number of sampling points $M$. As $M$ increases, the performance of both the proposed algorithm and the sequential update algorithm improves thanks to the increased position resolution. Notably, their performance gap becomes more significant compared to the single-user case, as shown in Fig.~\ref{sin_use}. This may be attributed to the more complex objective function in the multi-user case, which makes the sequential update algorithm more prone to getting trapped by undesired local optima. It is also observed that the performance of the PSO algorithm even degrades as $M$ increases, due to its insufficient exploration of feasible solutions. 
\begin{figure}[t]
\centerline{\includegraphics[width=0.4\textwidth]{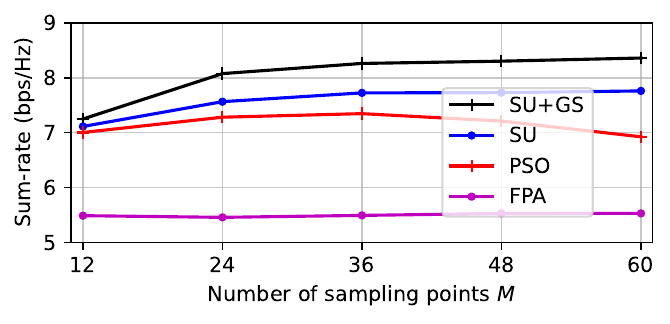}}
\vspace{-12pt}
\caption{Sum-rate versus the number of sampling points in multi-user case.}
\label{mul_use_m}
\vspace{-12pt}
\end{figure}

In Fig.~\ref{mul_use_lt}, we plot the sum rate versus the number of transmit paths $L_t$. It is observed that as $L_t$ increases, the performance of all schemes (except FPA) improves, thanks to the enhanced spatial diversity.  Moreover, the performance gap between the proposed algorithm and the sequential update algorithm increases with $L_t$, as the more complex propagation environment increases the possibility to achieve low-quality local optimality with the latter algorithm.
\begin{figure}[t]
\centerline{\includegraphics[width=0.4\textwidth]{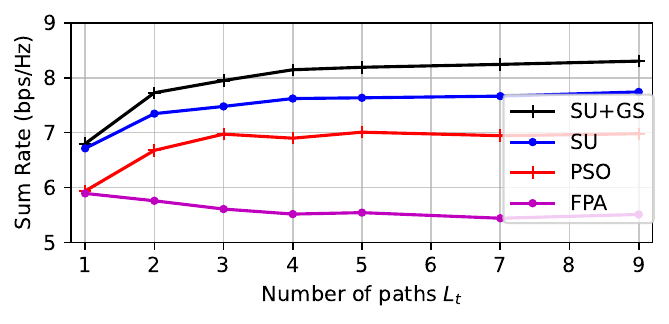}}
\vspace{-12pt}
\caption{Sum-rate versus number of transmit paths in multi-user case.}
\label{mul_use_lt}
\vspace{-12pt}
\end{figure}

In Fig.~\ref{mul_use_l}, we plot the sum rate versus the length of the transmit region, i.e., $A$, with the sampling resolution fixed as ${\lambda}/{8}$. As $A$ increases, the performance of all schemes with MAs is observed to improve, while that of the FPA benchmark remains constant. The PSO algorithm achieves a comparable performance to the sequential update algorithm given a sufficiently large value of $A$. However, both algorithms achieve a worse performance (around 0.8 bps/Hz) than the proposed algorithm.
\vspace{-6pt}

\begin{figure}[t]
\centerline{\includegraphics[width=0.4\textwidth]{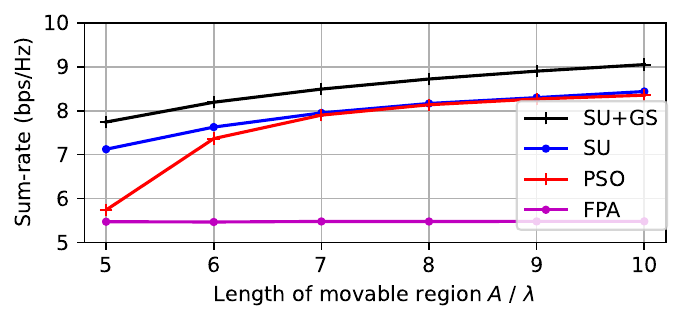}}
\vspace{-12pt}
\caption{Sum-rate versus length of movement region in multi-user case.}
\label{mul_use_l}
\vspace{-12pt}
\end{figure}

\section{Conclusion}
This letter proposed a general and low-complexity framework for antenna position optimization in MA systems. The proposed framework transforms conventional continuous position optimization into a discrete sampling point selection problem and solves it efficiently by combining the sequential update and GS. The GS was invoked between two consecutive rounds of sequential update to avoid low-quality local optima via solution exploration. Numerical results based on an MA-enhanced broadcast system showed that our proposed algorithm can achieve near-optimal performance and significantly outperform other benchmark schemes.
\vspace{-6pt}

\renewcommand{\baselinestretch}{1}
\bibliography{IEEEabrv, mybib.bib}

\begin{thebibliography}{10}
\providecommand{\url}[1]{#1}
\csname url@samestyle\endcsname
\providecommand{\newblock}{\relax}
\providecommand{\bibinfo}[2]{#2}
\providecommand{\BIBentrySTDinterwordspacing}{\spaceskip=0pt\relax}
\providecommand{\BIBentryALTinterwordstretchfactor}{4}
\providecommand{\BIBentryALTinterwordspacing}{\spaceskip=\fontdimen2\font plus
\BIBentryALTinterwordstretchfactor\fontdimen3\font minus
  \fontdimen4\font\relax}
\providecommand{\BIBforeignlanguage}[2]{{%
\expandafter\ifx\csname l@#1\endcsname\relax
\typeout{** WARNING: IEEEtran.bst: No hyphenation pattern has been}%
\typeout{** loaded for the language `#1'. Using the pattern for}%
\typeout{** the default language instead.}%
\else
\language=\csname l@#1\endcsname
\fi
#2}}
\providecommand{\BIBdecl}{\relax}
\BIBdecl

\bibitem{zhu2025tutorial}
L.~Zhu \emph{et~al.}, ``A tutorial on movable antennas for wireless networks,''
  \emph{{IEEE} Commun. Surveys Tuts.}, 2025, early access.

\bibitem{zhu2024modeling}
L.~Zhu, W.~Ma, and R.~Zhang, ``Modeling and performance analysis for movable
  antenna enabled wireless communications,'' \emph{{IEEE} Trans. Wireless
  Commun.}, vol.~23, no.~6, pp. 6234--6250, Nov. 2024.

\bibitem{mei2024movable}
W.~Mei, X.~Wei, B.~Ning, Z.~Chen, and R.~Zhang, ``Movable-antenna position
  optimization: A graph-based approach,'' \emph{{IEEE} Wirel. Commun. Lett.},
  vol.~13, no.~7, pp. 1853--1857, 2024.

\bibitem{zeng2025csi}
X.~Zeng \emph{et~al.}, ``{CSI}-free position optimization for movable antenna
  communication systems: A derivative-free optimization approach,''
  \emph{{IEEE} Wireless Commun. Lett.}, vol.~14, no.~1, pp. 53--57, Jan. 2025.

\bibitem{xiao2024multiuser}
Z.~Xiao \emph{et~al.}, ``Multiuser communications with movable-antenna base
  station: Joint antenna positioning, receive combining, and power control,''
  \emph{{IEEE} Trans. Wirel. Commun.}, vol.~23, no.~12, pp. 19\,744--19\,759,
  2024.

\bibitem{zhu2024multiuser}
L.~Zhu, W.~Ma, B.~Ning, and R.~Zhang, ``Movable-antenna enhanced multiuser
  communication via antenna position optimization,'' \emph{{IEEE} Trans.
  Wireless Commun.}, vol.~23, no.~7, pp. 7214--7229, Jul. 2024.

\bibitem{wei2024spectrum}
X.~Wei \emph{et~al.}, ``Joint beamforming and antenna position optimization for
  movable antenna-assisted spectrum sharing,'' \emph{{IEEE} Wireless Commun.
  Lett.}, vol.~13, no.~9, pp. 2502--2506, Sep. 2024.

\bibitem{wei2025movable}
X.~Wei, W.~Mei \emph{et~al.}, ``Movable antennas meet intelligent reflecting
  surface: Friends or foes?'' \emph{{IEEE} Trans. Commun.}, 2025, early access.

\bibitem{zhu2023movable}
L.~Zhu, W.~Ma, and R.~Zhang, ``Movable-antenna array enhanced beamforming:
  Achieving full array gain with null steering,'' \emph{{IEEE} Commun. Lett.},
  vol.~27, no.~12, pp. 3340--3344, Oct. 2023.

\bibitem{wang2025movable}
D.~Wang, W.~Mei, B.~Ning, Z.~Chen, and R.~Zhang, ``Movable antenna enhanced
  wide-beam coverage: Joint antenna position and beamforming optimization,''
  \emph{{IEEE} Trans. Wireless Commun.}, pp. 1--1, 2025.

\bibitem{lyu2025movable}
W.~Lyu, S.~Yang, Y.~Xiu, Z.~Zhang, C.~Assi, and C.~Yuen, ``Movable antenna
  enabled integrated sensing and communication,'' \emph{{IEEE} Trans. Wirel.
  Commun.}, vol.~24, no.~4, pp. 2862--2875, 2025.

\bibitem{shao2025exploit}
X.~Shao, R.~Zhang, and R.~Schober, ``Exploiting six-dimensional movable antenna
  for wireless sensing,'' \emph{{IEEE} Wireless Commun. Lett.}, vol.~14, no.~2,
  pp. 265--269, Feb. 2025.

\bibitem{hu2024secure}
G.~Hu \emph{et~al.}, ``Secure wireless communication via movable-antenna
  array,'' \emph{{IEEE} Signal Process. Lett.}, vol.~31, pp. 516--520, Jan.
  2024.

\bibitem{ding2025movable}
J.~Ding \emph{et~al.}, ``Movable antenna-aided secure full-duplex multi-user
  communications,'' \emph{{IEEE} Trans. Wirel. Commun.}, vol.~24, no.~3, pp.
  2389--2403, Mar. 2025.

\bibitem{ma2025robust}
H.~Ma \emph{et~al.}, ``Robust movable-antenna position optimization with
  imperfect {CSI} for {MISO} systems,'' \emph{{IEEE} Commun. Lett.}, vol.~29,
  no.~7, pp. 1594--1598, 2025.

\bibitem{mei2024secure}
W.~Mei, X.~Wei, Y.~Liu, B.~Ning, and Z.~Chen, ``Movable-antenna position
  optimization for physical-layer security via discrete sampling,'' in
  \emph{Proc. IEEE Global Commun. Conf.}, 2024, pp. 4914--4919.

\bibitem{wu2025globally}
Y.~Wu \emph{et~al.}, ``Globally optimal movable antenna-enabled multiuser
  communication: Discrete antenna positioning, power consumption, and imperfect
  {CSI},'' \emph{{IEEE} Trans. Commun.}, 2025, early access.

\bibitem{bremaud2013markov}
P.~Br{\'e}maud, \emph{Markov chains: Gibbs fields, Monte Carlo simulation, and
  queues}.\hskip 1em plus 0.5em minus 0.4em\relax Springer Science \& Business
  Media, 2013, vol.~31.

\end{thebibliography}
\bibliographystyle{IEEEtran}

\end{document}